\documentclass[10pt]{article}
\usepackage[OE]{express}

\usepackage{amsmath,amssymb,amsfonts}
\usepackage{graphicx}
\usepackage{comment}
\usepackage[normalem]{ulem}
\usepackage{lipsum,tabularx}
\usepackage{tabulary}
\usepackage{array}
\usepackage[title]{appendix}
\usepackage{epstopdf}
\usepackage{color}
\newcommand{\bra}[1]{\langle #1\vert}

\newcommand{\ket}[1]{\vert #1\rangle}

\newcommand{\pd}[1]{\frac{\partial}{\partial #1}}
\newcommand{\pdsq}[1]{\frac{\partial^2}{\partial #1^2}}
\newcommand{\refeq}[1]{Eq.~(\ref{#1})}

\newcommand{\reffig}[1]{Fig.~\ref{#1}}

\newcommand*\colvec[3][]{
    \begin{pmatrix}\ifx\relax#1\relax\else#1\\\fi#2\\#3\end{pmatrix}
}

 \newcommand{\polB}[0]{\mathbf{p}}
\newcommand{\muB}[0]{\boldsymbol{\mu}}
\newcommand{\angtwo}[0]{\Theta} 
\newcommand{\tensor}[1]{\boldsymbol{#1}}

 \newcommand{\rr}{\raggedright}
 \newcommand{\tn}{\tabularnewline}

\begin{document}

\title{Isolating the chiral contribution in optical two-dimensional chiral spectroscopy using linearly polarized light}

\author{David I. H. Holdaway,\authormark{1} Elisabetta Collini,\authormark{2} and Alexandra Olaya-Castro\authormark{1,*}}

\address{\authormark{1}Department of Physics and Astronomy, University College London, Gower Street, WC1E 6BT, London, United Kingdom\\
\authormark{2} Department of Chemical Sciences, University of Padova, via Marzolo 1, 35131 Padova, Italy
}

\email{\authormark{*}a.olaya@ucl.ac.uk} 

\begin{abstract}

The full development of mono- or multi-dimensional  time-resolved spectroscopy techniques incorporating optical activity signals has been strongly hampered by  the challenge of identifying the small chiral signals over the large achiral background.  Here we  propose  a  new methodology  to  isolate chiral signals removing the achiral background from two commonly used configurations for performing two dimensional optical spectroscopy, known as BOXCARS and GRadient Assisted Photon Echo Spectroscopy (GRAPES).  It is found that in both cases an achiral signal from an isotropic system can be completely eliminated by small manipulations of the  relative angles between  the linear  polarizations of the four input laser pulses. Starting from the formulation of a perturbative expansion of the signal in the angle between the beams and the propagation axis,  we derive analytic expressions that can be used to estimate how to change the polarization angles of the four pulses  to minimize achiral contributions in the studied configurations. The generalization to any other possible experimental configurations has also been discussed.
\end{abstract}

\ocis{(320.7150) Ultrafast spectroscopy; (190.4380) Nonlinear optics, four-wave mixing; (260.5430) Polarization; (050.1930) Dichroism.} 

\bibliographystyle{osajnl}

\section{Introduction}

Chirality is a structural property of systems lacking mirror symmetry. Almost all biomolecules are chiral (proteins, nucleic acids, sugars, etc.)  and many artificial materials currently employed in advanced photonic applications are also chiral (carbon nanotubes and graphene \cite{WongCarbon} and metamaterials \cite{Pendry1353,Ren2012}, for example).
The interest toward this kind of materials is justified by their peculiar optical properties, collectively defined as 'optical activity'. \cite{ParsonMOS}  The optical activity of a chiral sample strongly depends on  the intimate details of its molecular stucture and therefore it represents a fine tool to probe electronic and molecular structure.
For this reason, optical techniques, such as Circular Dichroism (CD) have been routinely employed to determine conformational and structural properties of these systems, for example in the assessment of secondary structures of proteins and other biologically relevant molecules \cite{FasmanCDBio,Tinoco1987} as well as the non-symmetric arrangement of pigments in light harvesting complexes. \cite{BerovaCDPaA,Georgakopoulou2006, Hemelrijk1992, Buchel1997,Furumaki2012}
If one then consider that  biological and artificial systems typically undergo time-dependent ultrafast structural changes during physical or chemical processes, the possibility of including an additional time-dependent dimension in conventional CD would also open the possibility of following how such ultrafast conformational  dynamics affect  the mechanism of relevant electronic processes. Moreover, chiral signals originate from components of a system response function which are independent of the achiral ones. Hence these signals provide independent information, which can be used to better determine chromophore structures and remove ambiguity about the origins of signals observed in achiral  time-resolved experiments. 

Despite their recognized potential, the full development of ultrafast time-resolved chiral techniques and the two-dimensional (2D) analogous has been strongly limited by the fact that chiral signals are typically three to four orders of magnitude weaker than achiral contributions \cite{BerovaCDPaA,ParsonMOS}.  This is problematic experimentally, as any chiral contributions can be obscured by even small contributions from achiral signals. Moreover most of the methods rely on differential measures (i.e. obtained by determining  the difference between signals generated by using left- and right-handed light) and therefore are particularly prone to background noise fluctuations.

Very few examples of time-resolved  optical activity measures can be found in the literature. After the pioneering work of Kliger's group,  \cite{Lewis1992} time-dependent chiral signals have been used to investigate the dynamic of structural changes in molecules and biomolecules \cite{Hache2009},  and vibrational optical activity\cite{Rhee2012}. There has also been a great deal of theoretical interest in using full chiral two-dimensional spectroscopy to better resolve cross peaks~\cite{Voronine2007}, study vibrational optical activity~\cite{Rhee2009}, geometric fluctuations~\cite{Mann2014} and coherence beating~\cite{Holdaway2016}. 

The possibility  of extending chiral measurements not only in the ultrafast time domain but also to multidimensional techniques  has been recently explored. The implementation of chiral sensitive schemes in 2D spectroscopy allowed characterizing the exciton delocalization following photoexcitation, and the evolution of coherent superpositions involving states with negligible transition dipole moments.\cite{FidlerEngel2014} Chiral 2D techniques could potentially  merge together the sensitivity of CD  measures to structural changes, with the capability of 2D spectroscopy to detect coherent dynamics in energy migration, unveiling important details about how structural motions  affect transport phenomena in biological  or artificial multi-chromophores systems.
A wider application of chiral 2D techniques require however the development of  experimental schemes able to efficiently isolate chiral signal from the  more intense achiral contribution, thus enhancing measurements sensitivity and  robustness against background noise.

Chiral signals can (in theory) always be isolated by using left and right circularly polarized pulses and comparing the two signals, or by numerically subtracting a known achiral component.  However, circularly polarized light is experimentally more challenging, and numerical subtraction requires noise levels below that of the final signal. Linear polarization control is simpler and is present in many existing 2D electronic spectroscopy (2DES) setups, it is used within non-chiral experiments to isolate off-diagonal and coherence contributions\cite{SchlauCohen2011,Lim2015}.  It is therefore useful to find ways to remove the non-chiral contributions at the single shot level (requiring no subtractions of weak signals) and requiring only linear polarization, which can be used in actual 2DES configurations.

In this work we derive sets of polarizations for which the achiral contributions to 2DES signals will cancel exactly in experimental configurations which are not co-planar.  The exact polarizations required depend on whether they are implemented in the BOXCARS~\cite{Fuller2015} or GRAPES~\cite{Harel2010,Harel2011} (GRadient Assisted Photon Echo Spectroscopy) configurations. The BOXCARS setup guarantees full phase matching and is the more commonly used configuration.  GRAPES manipulates a phase mismatch to allow single shot measurements, reducing the impact of amplitude noise in lasers\cite{Fuller2015}.


This work is divided into three sections. Section \ref{sec:Iso_av} describes the interaction with light and matter in the dipole and higher order multipole approximations and outlines the orientation averages which must be minimized, within the constraints set by our two chosen experimental geometries. Section \ref{sec:Canceling} derives analytic approximations for solutions in each case, which achieve cancellations of the achiral signal with minimal shifts to the polarizations compared with a 
fully parallel configuration. Finally section \ref{sec:Conclusion} summarizes our results and conclusions.

\section{Effect of isotropic averages to the chiral and achiral signal}
\label{sec:Iso_av}
\subsection{Multipole expansion of the Light-matter interaction}
The interaction of matter with electromagnetic radiation is usually treated with a multipole expansion of electron positions within the system of interest. In the case of a multichromophore aggregate as it is the interest of this paper, the small parameter in this expansion is $a/\lambda$ with $\lambda$  optical wavelength and $a$ the molecular size given by the displacement between chromophores~\cite{MukamelPoNLS}. The lowest order in this expansion is the electric dipole approximation, with electric dipole moment operator 
\begin{equation}
\hat{\boldsymbol{\mu}}(\mathbf{R}) = \sum_{\alpha} C_{\alpha}(\hat{\mathbf{r}}_{\alpha} - \mathbf{R})  \; ,
\end{equation}
with $\alpha$ running over all the component charges. Here $C_{\alpha}$ denotes the charge and $\hat{\mathbf{r}}_{\alpha}$ is the position operator for each charged particle. Higher order terms~\cite{Abramavicius2006} include the magnetic dipole $\hat{\mathbf{m}}$ and electric quadrupole $\tensor{Q}$ operators which are given by
\begin{align}
\hat{\mathbf{m}}(\mathbf{R})  = \sum_{\alpha} \frac{C_{\alpha}}{2m_{\alpha}c }(\hat{\mathbf{r}}_{\alpha} - \mathbf{R}) \times \hat{\mathbf{q}}_{\alpha}  \nonumber \; , \\
\hat{\tensor{Q}}(\mathbf{R})  = \sum_{\alpha} \frac{C_{\alpha}}{2 }(\hat{\mathbf{r}}_{\alpha} - \mathbf{R}) \times (\hat{\mathbf{r}}_{\alpha} - \mathbf{R}) \; . 
\end{align}
Here $\hat{\mathbf{q}}_{\alpha}$ denotes a momentum operator.  We outline the derivation of these terms in Appendix~\ref{SI:trans_current}. 

These terms can be combined into an interaction Hamiltonian~\cite{Cho2003}
\begin{equation}
H_I = -\hat{\muB} \cdot \mathbf{E}(\mathbf{r},t) - [\hat{\mathbf{m}} \cdot \mathbf{B}(\mathbf{r},t) + \tensor{Q} : \nabla \mathbf{E}(\mathbf{r},t)] \;,
\label{Eq:HI}
\end{equation}
here $\mathbf{E}$ and $\mathbf{B}$ are the electric and magnetic fields.  The notation $\tensor{Q} : \nabla \mathbf{E}$ denotes a tensor contraction over two indices, expressed in component parts that is $\sum_{j,k} Q^{j k} \nabla_{k} E_{j}$, with $j,k$ running over all spatial dimensions $x,y$ and $y$ and $Q^{xx}$ being the $x$ component of the tensor operator $\tensor{Q}$.

In 2DES experiments we detect the response of an ensemble, rather than signals from single molecules/complexes. Within this context, the purpose of this paper is to find the set of polarizations for which these ensemble averages will vanish unless they contribute a chiral signal.  
The dipole and quadrupole moments in \refeq{Eq:HI} are, respectively, vector and tensor quantities which do not vary significantly on the timescales of ultrafast spectroscopy. 
However, due to the random orientation of molecules in  isotropic solutions, all the dipole moments of a given system will be rotated randomly. The signals measured will correspond to that of an ensemble averaged over all molecular orientations. 

\subsection{Heterodyne signal detection in Nonlinear spectroscopy}
Within a nonlinear spectroscopy experiment, the medium, which we describe in terms of a density matrix $\hat{\rho}(\mathbf{r},t)$ is excited by a series of coherent laser pulses.  The medium then generates a new electric field through its own polarization, which is equal to $\rm{Tr}\{\hat{\boldsymbol{\mu}} \rho(t)\}$ within the dipole approximation. The dipole approximation is generally sufficient for the achiral signal component. To calculate chiral contributions, we must also include the magnetization and the quadrupole contributions to the polarization.

We can expand the time dependent density matrix perturbatively in $H_I(t)$ (see for example \cite{MukamelPoNLS}), giving $\rho(t) \sim \rho^{(0)}(t) +\rho^{(1)}(t) +\rho^{(2)}(t)+ \ldots$.  The $n$th order term produces the $n$th order polarization $\mathbf{P}^{(n)}(\mathbf{r},t)=\sum_j \mathbf{P}_j^{(n)}(t) \exp(i \mathbf{k}_{\mathrm{out},j}\cdot \mathbf{r})$. Two dimensional spectroscopy is a third order technique and hence we have $n=3$. 
Lower order terms do not contribute to the signal and higher orders are assumed to be negligible.  Assuming our pulses are all comprised of single wavevectors $\mathbf{k}_j$, then we measure the components with wavevectors $\mathbf{k}_{\mathrm{out},1} = -\mathbf{k}_1 + \mathbf{k}_2 + \mathbf{k}_3$, $\mathbf{k}_{\mathrm{out},2} = \mathbf{k}_1 - \mathbf{k}_2 + \mathbf{k}_3$ and $\mathbf{k}_{\mathrm{out},3} = \mathbf{k}_1 + \mathbf{k}_2 - \mathbf{k}_3$  corresponding to the rephasing, nonrephasing and coherence contributions. For completness, Appendix~\ref{SI:2DES} includes a note expanding on the principles of 2DES.

As we mentioned before, our macroscopic polarization is proportional to an ensemble average of all molecular orientations rather than a single molecule.  Since the polarizations of our electric fields are fixed, to compute $\mathbf{P}^{(n)}$ we must consider $n+1$th order tensor averages of our dipole moment operators, which we will discuss in Sec.~\ref{sec:FOIA}.

Our final signal is obtained by heterodyne detection with another electric field $\mathbf{E}_{n+1}(\mathbf{r},t)$ from a "local oscillator" (LO) pulse, hence 
\begin{equation}
S_j \propto \int_{-\infty}^{\infty} dt \; \mathrm{Im}[\mathbf{P}_j^{(n)}(t) \cdot \mathbf{E}^*_{n+1}(t)].
\label{eq:Hetdet}
\end{equation}
The signal depends on the scalar product of the polarization and LO field, and hence we can control which element(s) of the polarization field we actually measure.  When only parallel linear polarizations are used for the electric field which provide the $n$ interactions, say along the $x$ axis, an isotropic achiral medium produce a polarization only along the $x$ axis. However, chiral media can produce a nonlinear polarization with a component orthogonal to these field (say, along the $y$ axis). If the LO is polarized along $x$, then this "chiral" contribution will be lost and only the achiral contribution measured.  Where as, if the LO is polarized along $y$ instead, only the "chiral" contribution is measured. Notice also that the phase of the local oscillator relative to the other pulses is also used for selecting the real and imaginary parts of the signal field.

The chiral contribution is present regardless of whether we measure it. This therefore makes the polarization of the LO a valid parameter to consider in our analysis.  We should note that \refeq{eq:Hetdet} neglects terms proportional to $|P^{(n)}|^2$, as the induced signal field is typically far weaker than LO electric field.  This assumption is still expected to be valid when we are selecting the (much weaker) part of the signal field which is present only due the chirality of the medium.  


\subsection{Fourth order isotropic averages}
\label{sec:FOIA}
The achiral component of our signal in 2DES is proportional to $\rm{Tr}(\hat{\boldsymbol{\mu}} \rho^{(3)}(t))$, averaged over all global orientations of our system. To make evaluating these rotation averages easier, we can expand the electric dipole operator into components for each dipole allowed transition

\begin{align}
\hat{\muB} = \sum_{k} \muB_{k}\left ( \ket{0_k} \bra{k} + \ket{k} \bra{0_k} \right ) + \sum_{k \neq k'} \muB_{k'} \left ( \ket{k} \bra{k,k'} + \ket{k',k} \bra{k} \right )   \;, 
\end{align}
where $\ket{0_k}$ and $\ket{k}$ denote, respectively, the ground and excited states of  the $k-$th chromophore of our system and $\ket{k',k}$ denote two-exciton states. 

Analogous expansions can be presented for the magnetic dipole and electric quadrupole moment operators. Within this expansion, all terms which contribute to the achiral (electric dipole transitions only) signal can be broken down into terms which only depend on the internal dynamics of the system, with prefactors which are isotropic averages of the form

\begin{align}
\langle &(\mathbf{E}_1 \cdot \muB_{k_1}) \ldots (\mathbf{E}_{n+1} \cdot \muB_{k_{n+1}})\rangle_{\rm av}= \int_0^{2 \pi} d\alpha   \int_0^{\pi} d\beta \int_0^{2 \pi} d \gamma  \nonumber \\
&\frac{n \sin(\beta)}{8 \pi^2} (\mathbf{E}_1 \cdot T(\alpha, \beta, \gamma) \muB_{k_1}) \ldots (\mathbf{E}_{n+1} \cdot T(\alpha, \beta, \gamma) \muB_{k_{n+1}}).
\label{eq:isotropic_av_prefactor}
\end{align}
Only $n=3$ terms contribute significantly for a third order technique such as 2DES. Analogous expressions can be derived for the terms contributing to the chiral component. In that case the electric dipoles in Eq.(\ref{eq:isotropic_av_prefactor}) are  replaced with a magnetic dipole or electric quadrupole moment. The rotation matrix $T(\alpha, \beta, \gamma)$ is defined as~\cite{Wagniere1982}
\begin{equation}
T(\alpha, \beta, \gamma) = \begin{bmatrix}
 c_1 c_2 c_3 - s_1 s_3 &  - c_3 s_1 - c_1 c_2 s_3 & c_1 s_2 \\
 c_1 s_3 + c_2 c_3 s_1 & c_1 c_3 - c_2 s_1 s_3 & s_1 s_2 \\
 - c_3 s_2 & s_2 s_3 & c_2 \;,
\end{bmatrix}
\label{eq:Rot_mat}
\end{equation}
with $s_1 = \sin(\alpha)$, $s_2 = \sin(\beta)$, $s_3 = \sin(\gamma)$ and $c_1 = \cos(\alpha)$, $c_2 = \cos(\beta)$, $c_3 = \cos(\gamma)$. 
Integrals such as those in \refeq{eq:isotropic_av_prefactor} are known analytically, and hence we can consider:~\cite{Cho2003}
\begin{equation}
\langle \muB_1 \cdot \polB_1 \; \muB_2 \cdot  \polB_2  \; \muB_3 \cdot  \polB_3 \; \muB_4 \cdot  \polB_4 \rangle_{\rm iso} = \Xi \cdot \colvec[(\muB_1 \cdot \muB_2)(\muB_3 \cdot \muB_4)]{(\muB_1 \cdot \muB_3)(\muB_2 \cdot \muB_4)}{(\muB_3 \cdot \muB_2)(\muB_1 \cdot \muB_4)} \;.
\label{eq:4th_rank_av}
\end{equation}
For compactness we have written $\muB_j \equiv \mu_{k_j}$ and $\polB_j$ represent the polarizations of the fields responsible for the $j$th interaction.  Additionally we have
\begin{equation}
\Xi = \frac{1}{30}
\left( \begin{array}{ccc}
4 & -1 & -1 \\
-1 & 4 & -1 \\
-1 & -1 & 4 \end{array} \right) \colvec[(\polB_1 \cdot \polB_2)(\polB_3 \cdot \polB_4)]{(\polB_1 \cdot \polB_3)(\polB_2 \cdot \polB_4)}{(\polB_1 \cdot \polB_4)(\polB_2 \cdot \polB_3)} \;,
\label{eq:Xi}
\end{equation}
a three component vector, each element of which controls the amount each linearly independent component of $(\muB_1 \cdot \muB_3)(\muB_2 \cdot \muB_4)$ contributes to the signal using the chosen set of polarizations. The function $\Xi$ is related to the inner products of all possible pairs of polarizations $\polB_j$, which are unit vectors and hence is related to the angles between the different polarizations, which we denote 
\begin{equation}
\alpha_{j,k} = \cos^{-1}(\polB_j \cdot \polB_k)\;.
\end{equation}

The system response is contained purely in the final term on the right hand side of \refeq{eq:4th_rank_av}, and there is no reason to assume any component will vanish.
Therefore, the only way to guarantee these averages, and hence the achiral signal, are zero is if all components of $\Xi$ are zero. If this is the case our signal will   have no achiral component and the chiral contribution will be fully isolated. Our objective is then to find the polarization combinations for which ${\Xi}=\vec{0}$ without canceling out any chiral response. There are many possible solutions which achieve this. For example, we can take $\alpha_{1,2} = \alpha_{1,3} = \alpha_{1,4} =\pi/2$ along with any arbitrary values for the remaining angles $\alpha_{2,3}, \alpha_{2,4}$ and $\alpha_{3,4}$.  This is however not quite as simple as it first appears, because polarizations cannot be chosen independently of the direction of propagation. We discuss the best way to obtain these solutions in Sec.~\ref{sec:constraints}. For convenience we define the quantity
\begin{equation}
\tilde{\Xi} = |\Xi(1)|+|\Xi(2)|+|\Xi(3)| \;,
\end{equation}
where $|\Xi(j)|$ are the norms of each element of the length three vector $\Xi$. Clearly $\tilde{\Xi}=0$ if and only if $\Xi=\vec{0}$.

\subsection{Constraints on the polarizations of pulses}
\label{sec:constraints}

As light is a transverse wave in neutrally charged media, a single mode solution must have the electric field orthogonal to the direction of propagation and the magnetic field is then defined via $\mathbf{B} = \mathbf{k} \times \mathbf{E}/\omega$.  We assume each pulse consists of light propagating along a single direction $\mathbf{k}_j$ and must therefore have a polarization $\polB_j$ satisfying $\polB_j \cdot \mathbf{k}_j = 0$. The index $j$ on $\polB_j$ runs from $1$ to $4$, with $1$ to $3$ being the polarizations of the pulses which interact with the sample in time order and $4$ being the polarization of the local oscillator. 
Chiral signals are usually obtained using circular polarized (CP) light. The experiment is performed twice with the first pulse pulse being left CP and the second one right CP. The two signals are then subtracted from one another to reveal the CD signal. Such a differential spectroscopy can be performed in any experimental geometry, regardless of the angles between the wavevectors of each pulse. However, linearly polarized (LP) pulses are simpler to implement experimentally, hence we aim to determine how to perform chiral sensitive 2DES using only LP light. As above-mentioned, this is highly non-trivial as the condition $\polB_j \cdot \mathbf{k}_j  = 0$ must be satisfied. In order to formulate this constraint mathematically, we construct all possible linear polarization configurations by considering only rotations of a beam traveling along $z$ which is polarized along $x$. 

These rotations can be described by a matrix $T(\theta_1,\angtwo,\theta_3)$ which depends on three angles.  We use the same definition \refeq{eq:Rot_mat}, that it is a rotation by $\theta_1$ around the $z$ axis, then $\angtwo$ around the $y$ axis and then $\theta_3$ once again around $z$. As the wavevector $\mathbf{k}$ is initially along the $z$ axis, the first rotation does not effect $\mathbf{k}$ but does change the polarization. Fixing $\mathbf{k}_1$ to  $\mathbf{k}_4$ to specific values leaves $\theta_1$ as the only free and unconstrained variational parameter for each pulse in a given configuration.

\begin{figure}[ht]
\centering\includegraphics[width=0.5\linewidth]{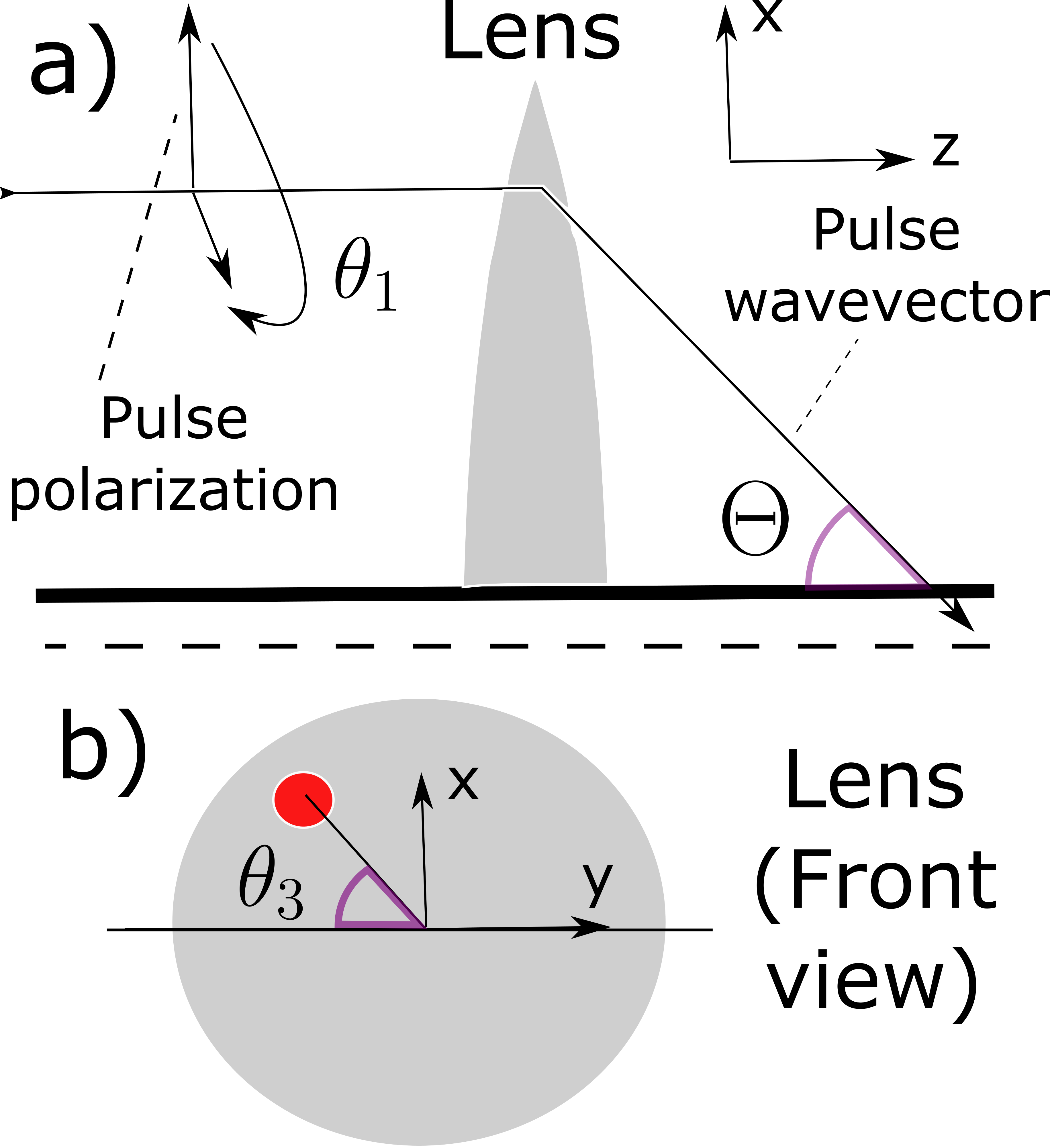}
\caption{Color online: Illustration of the physical meaning of the three rotation parameters in a lens based configuration.  a) Side view, b) Front view, the red circle shows entry point of the pulse.}
\label{Fig:lens_ray}
\end{figure}

Experimentally these rotations can be understood by considering a pulse traveling into a lens which focuses it into the centre of our sample. Assuming the pulse is initially traveling down the $z$-axis, we can interpret these angles as:  
\begin{enumerate}
  \item $\theta_1$ the angle of the polarization relative to $x$ before entering the lens.
  \item $\angtwo$ the angle between the direction of propagation into the sample and the principal axis of the lens. 
  \item $\theta_3$ is angle between the $y$-axis and a line drawn from the centre of the lens to the position point the pulse entered.  
\end{enumerate}
These angles are displayed in \reffig{Fig:lens_ray}.  The only angle which we assume experimental control of is $\theta_1$, while the others remain fixed. We denote the rotation angle of the $j$th pulse as $\theta_{1,j}$,  which constitute the four free parameters of interest.

\subsection{Minimisation of the achiral signal}
\label{sec:minimisation}
For our aim of measuring pure chiral signals, we  derive solutions in which the achiral component vanishes but the chiral contribution remains with high amplitude.  To achieve this, we set each component of the vector ${\Xi}$ to zero and find the pulse rotations $\theta_{1,1}$, $\theta_{1,2},\theta_{1,3}$ and $\theta_{1,4}$ for which this conditions is satisfied. This is in general not a simple problem as there can be many solutions for which these vector components will be zero i.e. we have four angles and only three equations to set to zero.  The simplest example to consider has all $\mathbf{k}_j$ along $z$ (colinear). We then have solutions whenever one of the four pulses is polarized orthogonal to the others (e.g.~$\polB_1$ along $y$ and all others along $x$ or $-x$), which gives four solutions plus a continuous degree of freedom, as we can make an arbitrary rotation about the $z$ axis on all $\polB_j$ without actually changing anything since the system is isotropic. Notice also that the minus factors in the polarizations are equivalent to a relative phase shift of $\pi$ to the other beams. These four configurations cancel all the achiral components in all the 2DES signal components and will all result in different, purely chiral signals.  However, only three of these signals are linearly independent of one another~\cite{Abramavicius2006}, that is, the signal obtained when $\polB_4$ is taken to be along $y$ (all others along $x$), can be constructed as a sum of the signals using $\polB_{j=1,2,3}$ along $y$ with all $\polB_{j'\neq j} = x$.

Polarizations schemes with one pulse polarization orthogonal to the other three can also be used within a pump-probe implementation using pulse shaping (c.f.~\cite{Zhang2015}) of the pump (with polarization control) and a Babinet-Soleil Compensator~\cite{Niezborala2007}. This is because the direction vectors of the pump pulse and probe pulse all lie in a common plane (corresponding to
the situation in which we set all $theta_3=0$ in \reffig{Fig:lens_ray}).  We can therefore choose one pulse to a have a polarization normal to this plane and others lying inside it. This geometry does allow for more independent signal components to be measured by changing $\angtwo$; we have shown in previous work that such a configuration can be used to study electronic coherence beatings~\cite{Holdaway2016}. In general, however, 2DES is not always performed in a totally colinear or coplanar configuration as it is not possible to separately measure the rephasing, non-rephasing and coherence contributions.  These contributions must instead be separated numerically via subtraction and phase-cycling~\cite{Tan2008}. This separation will be more technically demanding for chiral spectroscopy, as high amplitude achiral components would also need to be subtracted to leave a low amplitude chiral signal. 

Here we focus instead on configurations which are close to colinear, such as BOXCARS and GRAPES, and that have non-trivial solutions for which the achiral response vanishes. As these configurations can be seen as small deviations from colinear geometries we consider perturbatively changing from a colinear geometry, which leads to the rotation $\angtwo$ being the small parameter. In this way we can start with a solution to the colinear geometry ($\angtwo = 0$), for example $\polB_{1} = \hat{y}$ and $\polB_{2,3,4} = \hat{x}$, which we denote $\{yxxx\}$ and gradually change it to minimise the new $\tilde{\Xi}$, allowing us to keep track which of the possible colinear configuration this solution is closest to.  An exact analytic solution for $\Xi = \vec{0}$ (and thus $\tilde{\Xi} = 0$) has proved too difficult to obtain.  However, we have found it is possible to solve for arbitrary polynomial orders in $\angtwo^{2n}$. This is sufficient to describe any geometry as the series converges and $\Theta $ is an angle and therefore bounded. Essentially we perform this as follows: 
\begin{itemize}
 \item Start with solution to $\Xi = \vec{0}$ at $\angtwo = 0$, with all $\delta \theta_{j} = \theta_{1,j}(\Theta) - \theta_{1,j}(0)=0$ and $\theta_{1,j}(\Theta)$ determined by the experimental geometry.  
  \item Let $\delta \theta_j = A_j \angtwo^2$ and solve $\Xi = \vec{0}$ analytically in terms of $A_j$ along with the extra condition, $\sum_j \delta \theta_j = 0$, to remove the ambiguity in the solutions and keep them as close as possible to the unwrapped solutions.  
  \item  Repeat again with $\delta\theta_j = A_j \angtwo^2 + B_j \angtwo^4$ to solve for all the coefficients $B_j$ and so on for higher orders, until a sufficient order is achieved.
\end{itemize}
In order to construct the higher order solutions, we use the symbolic toolbox in MATLAB (equivalent to the MuPAD interface).  It would also be possible to derive the solutions by hand or using any other computer algebra program.

We note that as the medium is chiral, it can exhibit optical rotatory dispersion and cause the polarizations of the pulses to rotate and the angles between them will change.  We show in Appendix ~\ref{SI:CwsnT} that such effects will be negligible so long as the sample is not optically thick, which is a condition in general expected for non-linear spectroscopy.  


\section{Canceling the chiral signal in the BOXCARS and GRAPES setups.}
\label{sec:Canceling}
\subsection{Cancellation within the BOXCARS geometry}
\label{sec:boxcar}
\begin{figure}[ht]
\centering\includegraphics[width=0.7\linewidth]{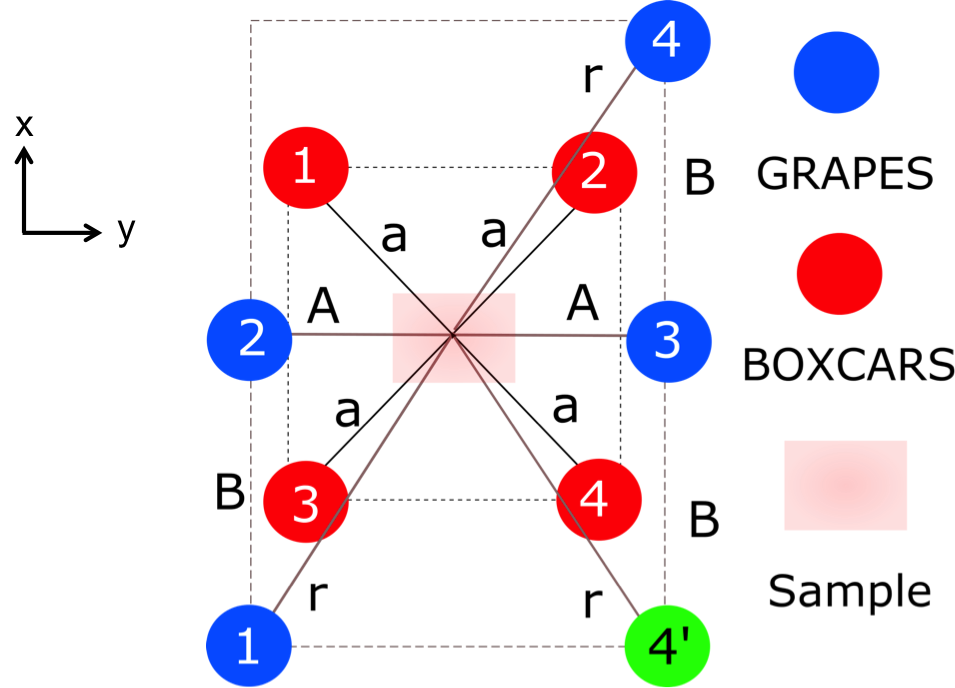}
\caption{Scheme of the spatial arrangement of the four pulses labeled from 1 to 4 in 2DES measures performed with BOXCARS (red dots) and rephasing/non rephasing GRAPES (blue/green) configurations when looking along the primary axis of propagation $z$. dashed lines indicate the square / rectangle in the $x-y$ plane identified by the four beams and a midpoint at the focal point having of the four beams within the sample. Solid lines show the components of $k_j$ orthogonal to $z$ and are labeled with the values of the absolute
distance to the centre point in the $x-y$ plane. }
\label{Fig:boxcar_GRAPES}
\end{figure}

We consider the effects of using the two beam configurations for 2D spectroscopy.  The first and the most common is the "BOXCARS" setup, shown as the red circles in \reffig{Fig:boxcar_GRAPES}, in which the directions of each of the pulses form a square pattern when one looks along the primary axis of propagation
\begin{equation}
\hat{k}_1 = \colvec[k_x]{-k_y}{k_z} \;, \quad \hat{k}_2 = \colvec[-k_x]{-k_y}{k_z}  \;, \quad \hat{k}_3 = \colvec[k_x]{k_y}{k_z} \;, \quad \hat{k}_4 = \colvec[-k_x]{k_y}{k_z} \;.
\end{equation}
This geometry ensures the phase matching condition is well satisfied and the separations between the pulse centers remain constant (assuming the refractive index of the media is approximately constant with frequency).  

Within the BOXCARS geometry, $\angtwo$ will be the same for all 4 beams and is determined by $\tan(\angtwo) = a / L$ with $L$ the distance from the central position to the focus in the sample and $a$ the displacement of the mirror (or lens) from the central position.  Finally assuming $0 \le \angtwo \le \pi$, we have $\theta_3 = 7\pi/4,5\pi/4,\pi/4,3\pi/4$ for beams $1,2,3,4$ respectively.  For convenience we introduce $\delta \theta_{j}=\theta_{1,j} (\angtwo) - \theta_{1,j} (+0)$ as the shift to the initial rotation at finite $\angtwo$ and $\delta \theta = \{ \delta \theta_{1},\ldots, \delta \theta_{4} \}$.  Physically this quantity is related to the rotation one needs to impose to the polarization optic used for each pulse

Within this geometry we can impose additional constraints due to the symmetry: $\delta \theta_{1} = -\delta \theta_{2}$ and $\delta \theta_{3} = -\delta \theta_{4}$, reducing the problem to a two parameter minimisation with a unique solution that keeps $\sum_j \delta \theta_{j} =0$.  
Adopting the nomenclature of~\cite{FidlerEngel2014}, two possible configurations will be investigated. First, we consider the 'chiral pump' configurations, in which transitions to a certain eigenstate involve a chiral interaction with one of the two pump pulses (pulse 1 or 2 in Fig.2). This is related to configurations with $\polB_1= \hat{y}$, or $\polB_2= \hat{y}$, with all others polarised along $\hat{x}$. These configurations are denoted $\{yxxx\}$ and $\{xyxx\}$, respectively. Second, we consider the 'chiral probe' configuration, denoted as $\{xxyx\}$, where for analogy $p_3=\hat{y}$ and all others are polarised along $\hat{x}$. For the first chiral pump configuration $\{yxxx\}$, at $\angtwo = 0$ we find $\theta_{1} = (\pi/4) [3,3,-1,-3]$. To order $\angtwo^4$ the expansion $\delta \theta_{1} = \angtwo^2/4 + C_1 \angtwo^4$ and $\delta \theta_{3}= 3\angtwo^2/4 + C_2 \angtwo^4$ minimises the achiral expression; the quartics coefficients are found to be $C_1 = 1/24$ and $C_2 = 15/24$.  

In the $\{xyxx\}$ configuration (other chiral pump) the shifts to $\delta \theta_j$ are exactly the same as the $\{yxxx\}$ case but now the initial rotation is given by $\theta_{1} = (\pi/4) [3,1,-1,-3]$.  
In the "chiral probe"  $\{xxyx\}$ configuration we have an initial rotation of $\theta_{1} = (\pi/4) [1,3,1,-3]$ and in terms of the values for the chiral pump shifts we have $\delta \theta_{1}^{\rm pr} =  \delta \theta_{4}$ and $\delta \theta_{3}^{\rm pr} = -\delta \theta_{1}$. These difference can be understood by looking at \reffig{Fig:boxcar_GRAPES} and noting that each is just given by a reflection of all the points in the vertical or horizontal planes.  
The results relating $\delta \theta_{j}^{\rm pr}$ to the shifts in the chiral pump configuration hold true for all higher order terms as well.

\begin{figure}[t]
\centering\includegraphics[width=0.9\linewidth]{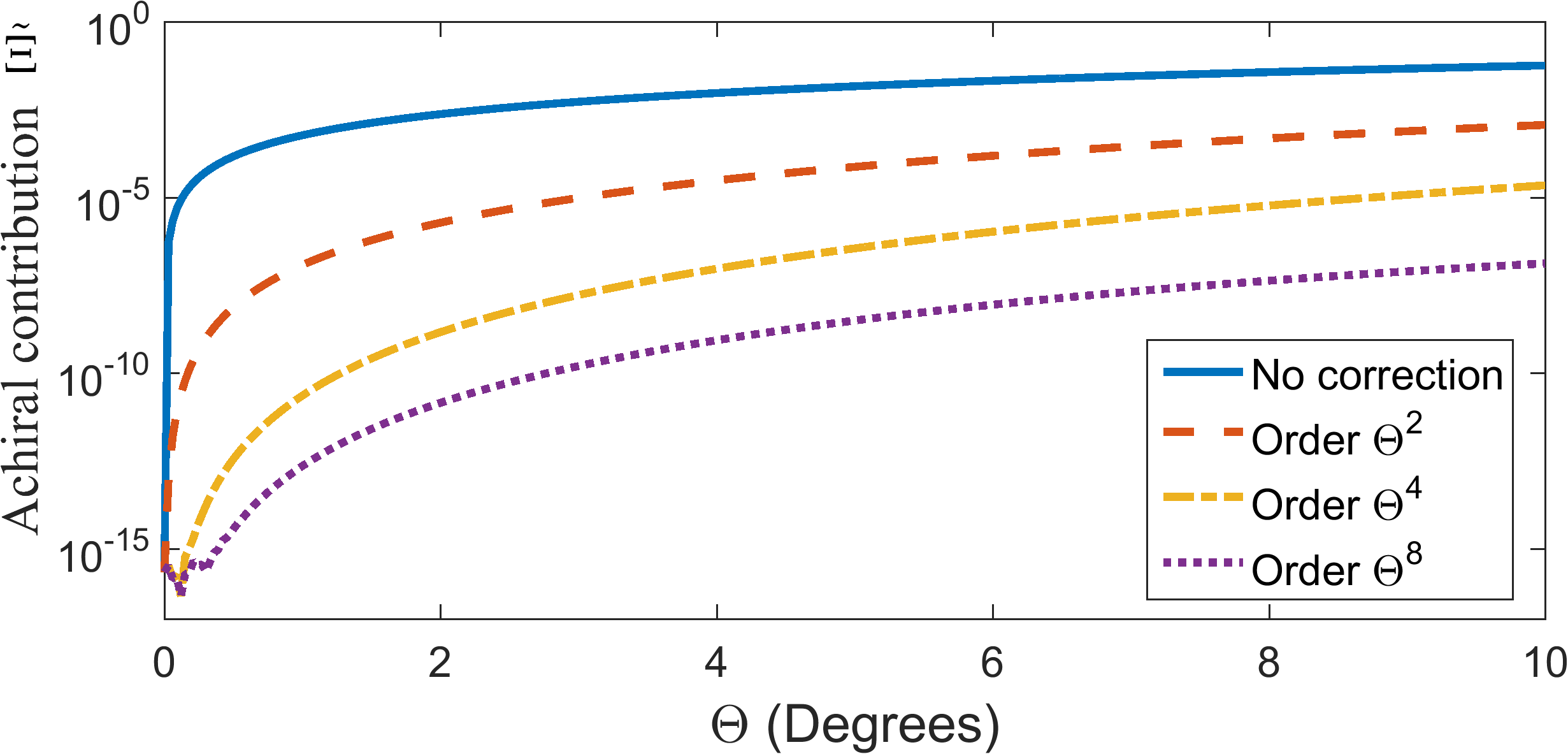} 
\caption{Amplitude of $\tilde{\Xi}$, the contribution of achiral signal, for a range of $\angtwo$ when correction different orders of correction are made. Note that 0.1 radians $\approx$ 5.73 degrees.  Noise in the graph below $10^{-15}$ are down to elementary precision errors.}
\label{Fig:boxcar_pol}
\end{figure}

In terms of the coordinate space polarizations, the changes up to quadratic order in $\angtwo$ for the $\{yxxx\}$ configuration are given in table \ref{table:delta_p} . Higher order corrections can be taken fairly easily from the expressions for the shifts to the angles.
\begin{table}[ht]
\centering
{\renewcommand{\arraystretch}{1.3}
    \begin{tabular}{| c | c | c | c |}
    \hline
    Pulse & $p_x$ & $p_y$ & $p_z$ \\ \hline
    1 & 0 & 1 $- \angtwo^2/4$ & $\angtwo/\sqrt{2}$ \\ \hline
    2 & $1 - \angtwo^2/4$ & $-\angtwo^2/2$ & $\angtwo/\sqrt{2}$ \\ \hline
    3 & $1 - \angtwo^2/4$ & $\angtwo^2/2$ & -$\angtwo/\sqrt{2}$  \\ \hline
    4 & $1 - \angtwo^2/4$ & $-\angtwo^2/2$ & $\angtwo/\sqrt{2}$ \\ 
    \hline 
    \end{tabular}}
   \caption{Chiral pump shifts $\delta\theta_j$ removing achiral signal contributions to order $\angtwo^2$  within the BOXCARS geometry, for the polarization configuration $\{yxxx\}$.} 
   \label{table:delta_p}
\end{table}
It is also important to investigate the changes to the chiral signal which are caused by the adjustment of these polarizations in order to interpret the signal.  We list the factors $(\mathbf{b}_j \cdot \polB_k)(\polB_{\ell} \cdot \polB_{m})$ which occur in the orientation averages for the four terms with the magnetic interactions up to fourth order in $\angtwo$.   
\begin{table}[ht]
\centering
{\renewcommand{\arraystretch}{1.3} 
  \begin{tabular}{| c| c | c | c |c |}    
    \hline
 Comp & $\mathbf{b}_1$ & $\mathbf{b}_2$ & $\mathbf{b}_3$ & $\mathbf{b}_4$ \\ \hline
 
[1,2 ; 3,4] & $-1 + 2\angtwo^2 + \angtwo^4/3$& $1 - \angtwo^2 - (5\angtwo^4)/3 $&   0&    0\\ \hline

[1,3 ; 2,4] &   -1&   0&$ 1 - \angtwo^2 + \angtwo^4/3 $&   0 \\ \hline

[1,4 ; 2;3] & $-1 + 2\angtwo^2 + \angtwo^4/3 $&   0&    0& $1- 2\angtwo^2  + 2\angtwo^4/3 $ \\ \hline
\end{tabular}}
\caption{Factors $(\mathbf{b}_j \cdot \polB_k)(\polB_{\ell} \cdot \polB_{m})$ which occur in isotropic averages relevant to TDS with magnetic interactions, calculated to order $\angtwo^4$.}
 \label{table:chiral_weights}
\end{table}
The changes here amount to about one part in 100 when $\angtwo \approx 0.1$ which is likely small enough to ignore, but can easily be taken into account.  

If the angles between the beams $\angtwo = \arctan(a / L)$ is not small this polynomial expansion is insufficient. However, it still appears to be possible to find values for the polarizations which cause the
achiral response to vanish, while leaving the chiral response finite, in all cases.  Numerically we find a correction to $\delta \theta_1$ to be of the order of  $(\cdots) +(0.6840 \angtwo^6  - 1.7634 \angtwo^8   - 0.6270 \angtwo^{10}) \times 10^{-3}$ with the $(\cdots)$ denoting the lower order terms listed in table \ref{table:delta_p}. We also have a correction to $\delta \theta_3$ of the order of $(\cdots)+ 0.4146 \angtwo^6  +  0.2183 \angtwo^8  -0.3670 \angtwo^{10}$,  which are a factor of $10^3$ larger.

\subsubsection{Example results with a Dimer system}

\begin{figure}[ht]
\centering\includegraphics[width=0.95\linewidth]{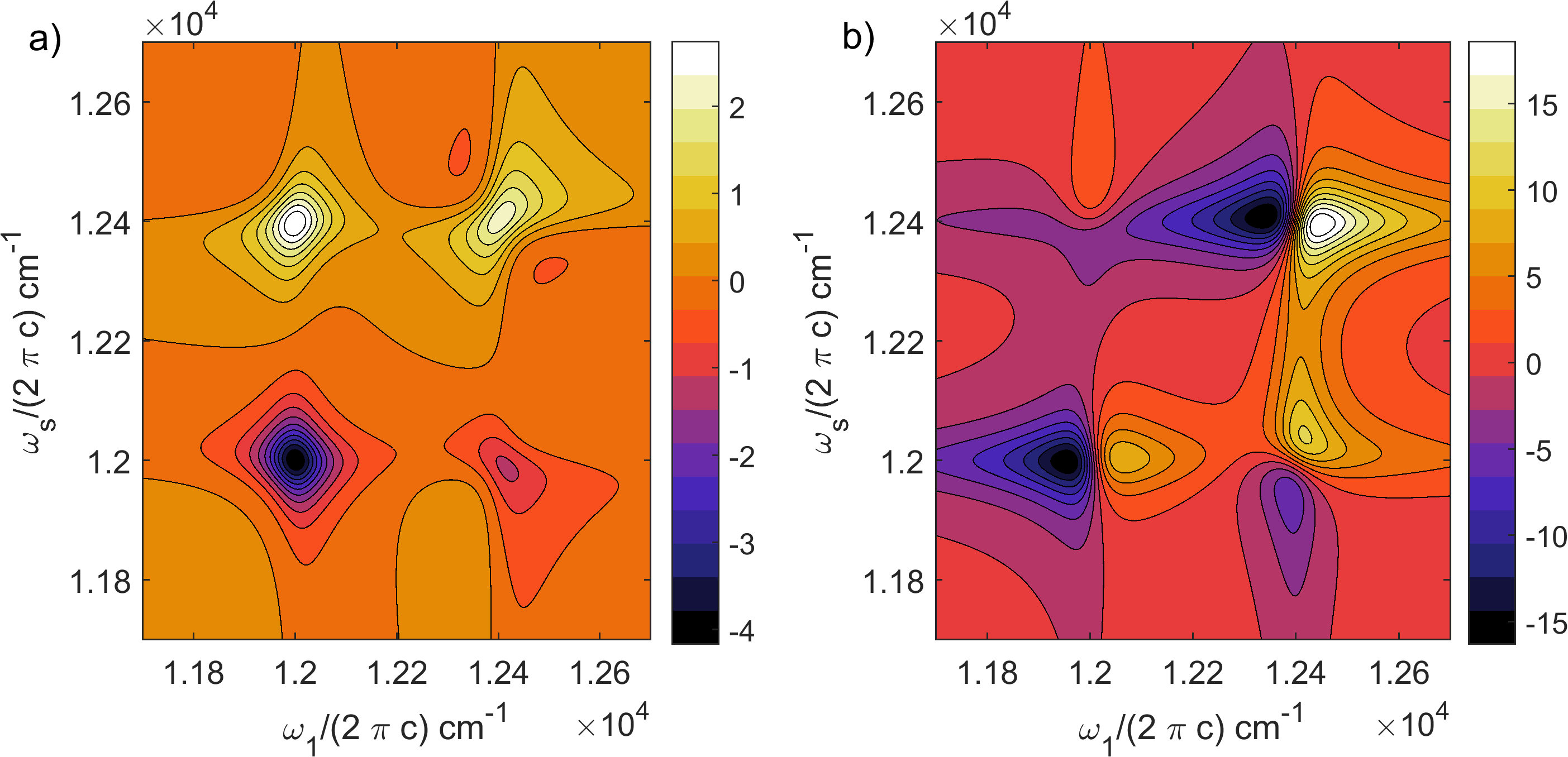} 
\caption{2D rephasing signal from an excitonic dimer (arbitrary units) in a BOXCARS configuration with $\angtwo = \pi/18$ (10 degrees) (a) Using the polarization shifts suggested in this work and (b) Without the shifts.  In (b) the signal has significant achiral contributions with $\tilde{\Xi} \approx 0.06$.}
\label{Fig:2D_sig}
\end{figure}

In \reffig{Fig:2D_sig} we plot theoretical data for a 2DES rephasing signal from a BOXCARS configuration at zero time delay for the simple dimer exciton system described in Appendix~\ref{SI:param}.  In \reffig{Fig:2D_sig}  a) we choose polarizations in the $\{xxyx\}$ chiral probe configuration with the correction included from having $\angtwo$ finite, where as b) has all $\delta\theta_j = 0$.  It is clear that the achiral contribution (dispersive rather than absorptive due to the $\pi/2$ phase difference between the local oscillator and the other pulses) is an order of magnitude larger, masking the chiral signal.  Typically $\angtwo$ is of the order of 10 degrees in experiments ($\pi/18$ radians), hence these effects are significant and cannot be ignored. 

\subsection{Cancellation within the GRAPES configuration}
Within the GRadient Assisted Photon Echo Spectroscopy (GRAPES) setup there is not the  clear cylindrical symmetry that is present in the BOXCARS setup and hence we require two parameters to describe the setup fully.  They can however be constructed in the same way using the three angles, except that beams 1 and 4 now have a different (larger) $\angtwo$ value to 2 and 3. We denote $\tan(\tilde{\angtwo}) = A/L$ the angle required for pulses two and three and $\tan(\angtwo) = r/L$ the angle for pulses 1 and 4, with $A$, $B$ and $r = \sqrt{A^2+B^2}$ as shown on \reffig{Fig:boxcar_GRAPES}.  We then have $\theta_{3} = 0,\pi$ for pulses 2 and 3 while $\theta_{3} = -\phi,-\phi+\pi$ with $\phi \equiv \arctan(B/A)$.  

\begin{figure}[ht]
\centering\includegraphics[width=0.95\linewidth]{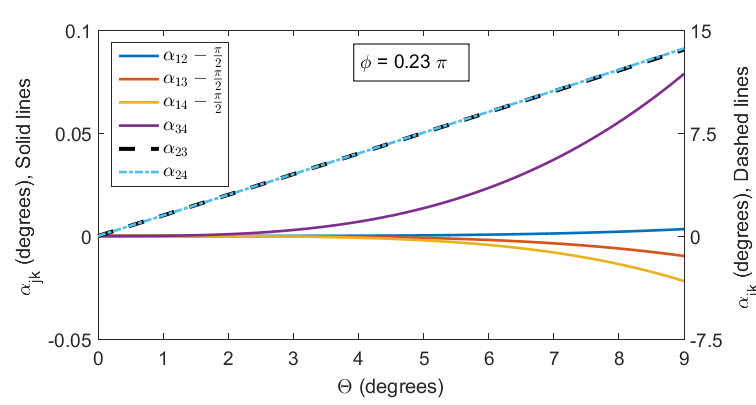}
\caption{Changes to angles between the four polarizations in the deformed $\{yxxx\}$ GRAPES configuration, for $\phi = 23 \pi/100$ and a range of $\angtwo$, using the 4th order approximation. The small changes to the angles between $\polB_1$ and the others three polarizations along with $\alpha_{34}$ are plotted on the left scale, where as $\alpha_{23}$ and $\alpha_{24}$ are plotted on the right scale. }
\label{Fig:angles_grapes}
\end{figure}


To get the same $\{yxxx\}$ geometry previously considered we can have $\theta_1 = \{\pi/2 + \phi,0,\pi,\pi+\theta \}$ when $r= 0$, for each of the beams 1 to 4.  As we increase $r$ we have shifts to these initial values of $\delta\theta =\{C+8,C+10,C+2,C\} \sin(2\phi)\angtwo^2/8$, with $C$ an arbitrary constant, which cancels the achiral contribution to second order in $\angtwo$, leaving 

\begin{equation}
\Xi = \frac{\angtwo^4}{192} \colvec[-(476\sin(2\phi) + 76\sin(4\phi) + 42C\sin(2\phi) + 27C\sin(4\phi))]{484\sin(2\phi) + 244\sin(4\phi) + 78C\sin(2\phi) + 33C\sin(4\phi)}{2(212\sin(2\phi) + 132\sin(4\phi) - 6C\sin(2\phi) + 39C\sin(4\phi))} + \mathcal{O}\left(\angtwo^6 \right) \;,
\end{equation} 
for the remaining signal.  
Similar solutions are also available for the $\{xyxx\}$ and $\{xxyx\}$ configurations with $\theta_1 = \{\phi,\pi/2,\pi,\phi\}$ and $\theta_1 = \{\phi,\pi,3\pi/2,\phi\}$ respectively and $\delta\theta =\{C,C+2,C+2,C\}$ in both cases. 

The polarizations of each beam using the minimisation for $\{yxxx\}$ are shown in table \ref{table:delta_p2}
with $C =-5$ taken to set $\sum_j \delta \theta_j = 0$ as we did in the BOXCARS geometry. The expression is however considerably more complicated due to the extra parameter $\phi$.


    

 


\begin{table}[t]
{\renewcommand{\arraystretch}{1.3}
\centering 
\begin{tabular}{| c |m{3.6cm} | m{3.6cm} |m{3.6cm} |}\hline

  Pulse & $p_x$ & $p_y$ & $p_z$ \\ \hline 
     
1 & \rr $\cos(U+\phi) \sin(\phi) - \sin(U+\phi) \cos(\angtwo) \cos(\phi)$ 
& \rr $ \cos(U+\phi) \cos(\phi) + \sin(U+\phi) \cos(\angtwo) \sin(\phi) $
& \rr $\sin(U+\phi) \sin(\angtwo)  $ \tn  \hline 
  
2 & \rr $\cos(V) \cos(\angtwo)$ & \rr $ \sin(V)$ & \rr 
$-\cos(V) \sin(\angtwo)$ \tn \hline
 
3& \rr $\cos(U) \cos(\angtwo)$ 
 & \rr $-\sin(U)$ & 
\rr $ \cos(U) \sin(\angtwo)$ \tn \hline 

4& \rr $\sin(\phi - V) \sin(\phi) + \cos(\phi - V) \cos(\angtwo) \cos(\phi)$ & \rr $ \sin(\phi - V) \cos(\phi) - \cos(\phi - ) \cos(\angtwo) \sin(\phi)$ 
 & \rr $ \cos(\phi - V) \sin(\angtwo)$\tn \hline
    \end{tabular}}
\caption{Chiral pump shifts $\delta\theta_j$ which remove achiral signal contributions up to order $\angtwo^2$, within the GRAPES geometry for the polarization configuration $\{yxxx\}$. For compactness, we have defined   $V \equiv 5 \angtwo^2  \sin(2 \phi) /8$ and $U \equiv 3 \angtwo^2  \sin(2 \phi) /8$, which are used only in this table.} 
\label{table:delta_p2}
\end{table}

We can also extend our expansion relatively easily to fourth order and obtain  $\delta\theta  = (\cdots) + \angtwo^4\{13\sin(2\phi)/64 + \sin(4\phi)/128,61\sin(2\phi)/192 - 23\sin(4\phi)/384,- 11\sin(2\phi)/192 - 31\sin(4\phi)/384,17\sin(4\phi)/128 - 89\sin(2\phi)/192 \} + \mathcal{O}(\angtwo^6)$, which eliminates all terms in the chiral response up to 6th order in $\angtwo$. When $\angtwo <\pi/18$ (or 10 degrees), $\tilde{\Xi}$ is less than $4 \times 10^{-6}$ at its maximum near $\phi \sim \pi/4$.  

In \reffig{Fig:angles_grapes} we show the angles between the polarizations for this minimization scheme (when $\phi = 0.23 \pi$), the angles between the "chiral" first pulse (polarized along the y-axis when $\angtwo=0$) and the other three pulses remains small, and hence the angles remain close to $\pi$. The angles $\alpha_{23}$ and $\alpha_{24}$ depend approximately linearly on $\angtwo$, where as change to the other angles is greater than quadratic. 

\subsubsection{Example results for a Dimer system}

\begin{figure}[tbp]
\centering\includegraphics[width=0.95\linewidth]{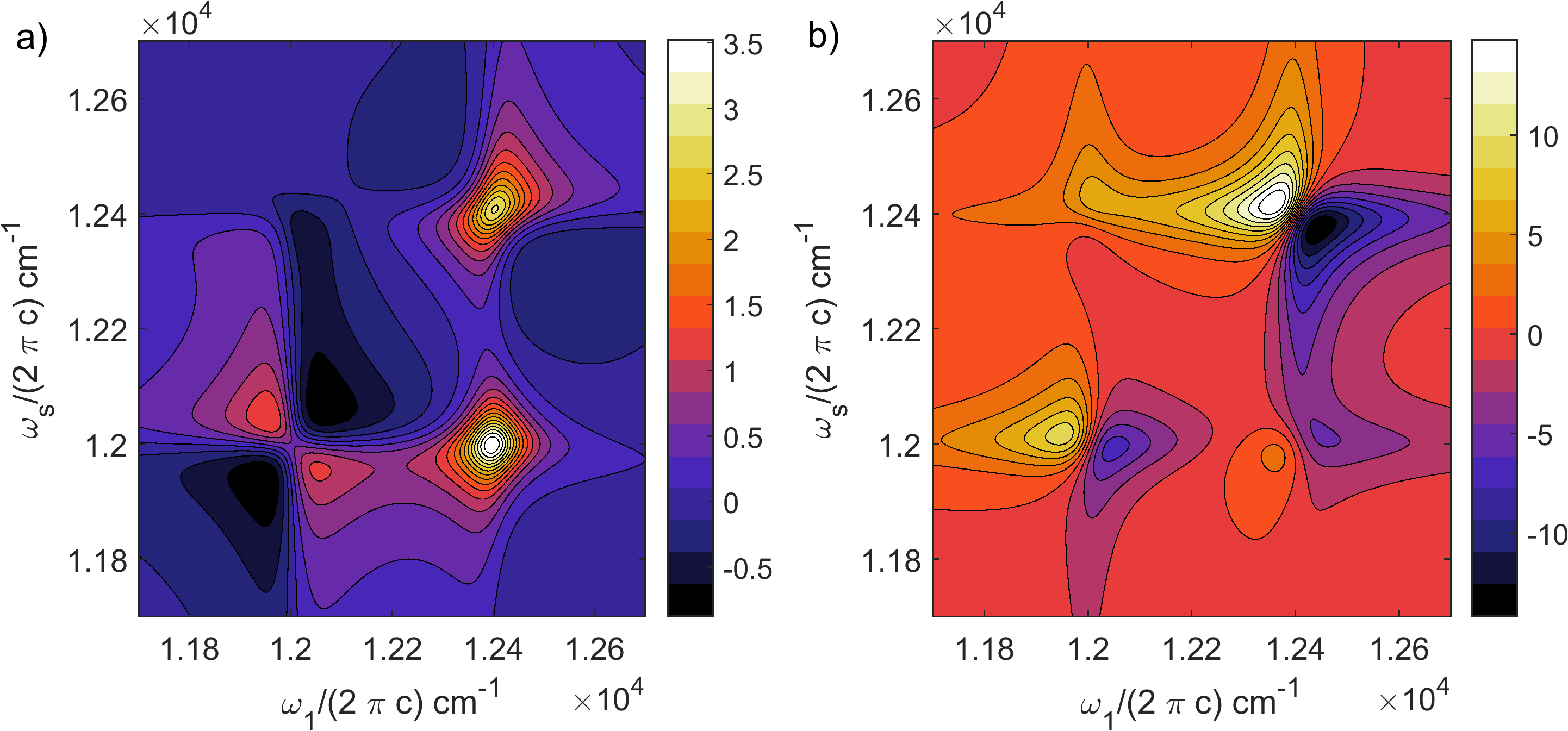}
\caption{2DES rephasing signal at zero population time, for an excitonic dimer in a GRAPES configuration with $\angtwo = \pi/18$ (10 degrees) and $\phi = 11\pi/90$ in a warped chiral pump $\{yxxx\}$ configuration. (a) Using the polarization shifts suggested in this work and (b) Without the shifts.  The chiral pump configuration gives a significantly different signal to the chiral probe shown before, providing further information about the system.} 
\label{Fig:2DS_grapes}
\end{figure}

We again generate theoretical results for the model dimer system described in the Appendix \ref{SI:param}, but using the GRAPES configuration instead.   In \reffig{Fig:2DS_grapes} we show simulated 2DES signals (at zero population time) with (a) and without (b) the polarization shifts, within a chiral pump (warped $\{yxxx\}$) configuration.  The overall signals in the pure chiral signal are quite significantly different from \reffig{Fig:2D_sig} as a result of using the different configuration. Excluding contributions from coherent dynamics, only the interaction with the first pulse will include a magnetic dipole.  This means that the magnetic dipole moments of transitions to the double excited states will never contribute; Feynman diagrams corresponding to transitions to double excited states will still contribute to the signal, but the chiral interactions must be the first ones.  Peaks associated with double excited state transitions at zero population time will be located off the lead diagonal. The changes to the two cross peaks in \reffig{Fig:2DS_grapes} a)  is clearly visible when compared to \reffig{Fig:2D_sig} a), with the upper left peak undergoing full cancellation.  Additionally new peaks appear, with opposite sign at different values of $\omega_1$ since the two excitons have opposite effective magnetic dipole moments.  
In b), again we see a strong contribution from the dispersive part of the achiral signal if the shifts are not performed, obscuring the chiral signal.  The amplitude of this achiral signal is roughly the same as that found in \reffig{Fig:2D_sig} b).

\subsection{Impact of errors in polarization alignment}

\begin{figure}[ht]
\centering\includegraphics[width=0.95\linewidth]{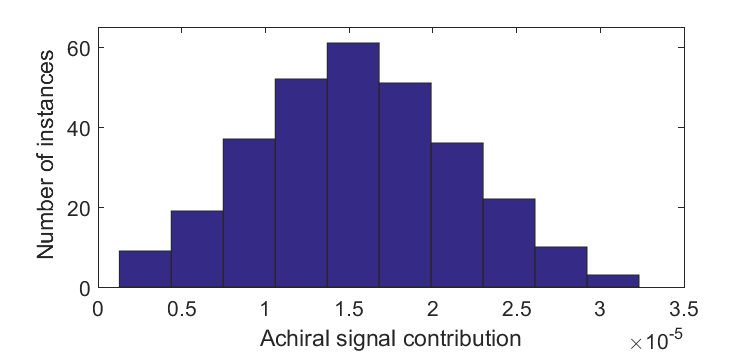}
\caption{Histogram of the amplitude of the achiral contribute, $\tilde{\Theta}$, to our "chiral pump" GRAPES configuration at $\angtwo = 10$ degree, with all $\delta\theta_{j}$ having uniform random noise added with an amplitude of 1/120 degrees.}
\label{Fig:hist_error}
\end{figure}

Precision polarization rotation optics are rarely sensitive to less than one arcminute (1/60 of a degree).  For this reason it is important to consider the sensitivity of these achiral signal minimizing solutions to small errors in the polarization controls. 

Even with (uniform) random deviations of a single arcminute to the polarization rotations away from our ideal values, chiral contributions become significant.  In fig\ref{Fig:hist_error} we plot a histogram of the achiral contributions with random noise.  This contribution is around an order of magnitude lower than the peaks of the chiral contributions (scaled by a factor of $10^{4}$ in Fig.~\ref{Fig:2DS_grapes}) but is still fairly significant as a noise source, even at this high level of polarization control.  The distribution appears approximately Gaussian, and the mean is directly proportional to the average error in the small noise limit. 

With some calibration using an achiral sample, it should be possible to numerically subtract such a contribution (assuming regular achiral spectroscopy has been performed on the sample of interest). Unlike in other signal subtraction techniques, such as those using left/right CP light or pulse shaping, the achiral signal should have a comparable amplitude to the chiral signal. Therefore if this subtraction is required it should be less prone to noise sources.  Stability of polarization optics is key, as fluctuations would be difficult to mitigate.

\section{Conclusion}
\label{sec:Conclusion}
We have put forward a technique that allows full cancelation of the non-chiral signal and therefore isolate a pure chiral contribution to signals in typical 2DES  geometries of non-colinear experiments, namely the BOXCAR and GRAPES configurations.  Our scheme relies in modifying three colinear polarization schemes.
In a truly colinear/coplanar 2DES geometry, these three polarization schemes would result in three linearly-independent chiral signals.  We find that it is possible to derive polynomial expansions using the angles between the beams and the $z$ axis to make any achiral contribution arbitrarily small, such that it can be neglected to obtain a pure chiral signal. This is of signficant experimental relevance as achiral contributions can dominate the 2DES signals in typical configurations making difficult to measure the chiral response.

These modifications to the polarizations of the beams as the geometry is warped, also modify the chiral contribution to the signal. Therefore, signals relating to the other two linearly-independent colinear configurations will contribute, along with some non-colinear configurations. The changes are of order $\angtwo^2$, with $\angtwo$ the angles between the beams and the $z$ axis (the direction in which the nonlinear signal grows), and so these changes can be neglected when $\angtwo^2 \ll 1$.  If this is not the case, these changes would need to be considered in the analysis of data obtained from the BOXCARS or GRAPES configurations.  However, this would not present significant complications in terms of the analysis.

The precise control of the polarizations required is likely to be experimentally challenging, and experiments may need to test configurations using achiral samples to quantify the impact of polarization misalignment.  For experiments aiming to minimise achiral contributions through calibration, the achiral minimizing solutions provided in this paper and the techniques to derive them, will still provide a good starting point for any optimization of the polarizing optics.  

\section*{Funding}
We gratefully acknowledge financial support from the EU FP7 project PAPETS -- Phonon-Assisted Processes for Energy Transfer and Sensing (GA 323901); and the ERC Starting Grant QUENTRHEL (GA 278560).

\begin{appendices}
\section{Interactions beyond the dipole approximation}
\label{SI:trans_current}
The minimum coupling Hamiltonian for light and matter in the semi-classical approximation is given by~\cite{MukamelPoNLS}:
\begin{equation}
 \hat{H}'(t) = -\int d \mathbf{r} \left[ \hat{\mathbf{J}}(\mathbf{r},t) \cdot \mathbf{A}(\mathbf{r},t) + \hat{Q}(\mathbf{r},t) : \mathbf{A}(\mathbf{r},t) \mathbf{A}(\mathbf{r},t)\right] \; .
 \label{eq:min_coup_Ham_full}
\end{equation}
We neglect the term proportional to the square of the (classical) electromagnetic vector potential $\mathbf{A}(\mathbf{r},t)^2$ as contributions from this term are typically small compared with $\hat{\mathbf{J}}(\mathbf{r},t) \cdot \mathbf{A}(\mathbf{r},t)$. This assumption is valid when the pulses do not overlap significantly.  If the pulses do overlap, then $(\mathbf{A}_j + \mathbf{A}_k)^2$ will give rise to cross terms which are proportional to $\exp(\pm(\mathbf{k}_j-\mathbf{k}_k) \cdot \mathbf{r})$ multiplied by the overlap between the envelopes.  

We can then express the effective semi-classical Hamiltonian in $\mathbf{k}$ space as
\begin{equation}
 \hat{H}'(t) \approx -\int d \mathbf{k}  \hat{\mathbf{J}}(\mathbf{k},t) \cdot \mathbf{A}(-\mathbf{k},t) \; .
 \label{eq:min_coup_Ham_FT}
\end{equation}
Denoting the creation and annihilation operators for the $a$th excited state of the $\ell$th chromophore $\hat{B}^{\dagger}_{\ell a}$ and $\hat{B}_{\ell a}$, 
the current density operator in momentum space becomes 
\begin{equation}
 \hat{\mathbf{J}}(\mathbf{k},t) = \sum_{\ell,a} \left( \overline{j}^*_{\ell a}(-\mathbf{k}) \hat{B}^{\dagger}_{\ell a} + \overline{j}_{\ell a}(-\mathbf{k}) \hat{B}_{\ell a} \right) \;.
\end{equation}
The terms $\overline{j}_{\ell a}(\mathbf{k})$ can in principle be calculated from the many-body wavefunctions of the ground and excited states via a multipole expansion in the displacement of charges from the chromophore center~\cite{Abramavicius2006-2}:
 \begin{align}
 \overline{j}_{\ell a}(-\mathbf{k}) =  -&i e^{i \mathbf{k} \cdot \mathbf{r}_j} \sum_{\alpha} C_{\alpha} \bra{\phi_{\ell a}}   \omega \left[(\mathbf{r}_{\alpha}-\mathbf{r}_{j}) \right.  \nonumber \\
 &-i \left. \mathbf{k} \cdot (\mathbf{r}_{\alpha}-\mathbf{r}_{j}) \otimes (\mathbf{r}_{\alpha}-\mathbf{r}_{j})/2 + \ldots\right]  \nonumber \\ 
 + &\mathbf{k} \times \left[ (\mathbf{r}_{\alpha}-\mathbf{r}_{j}) \times \mathbf{p}_{\alpha}/2j_{\alpha} + \ldots \right]  \ket{\phi_{jg}}  \;.
\label{eq:full_trans_current}  
\end{align}
Note that $c=\hbar =1$ in the above expression and the "$\ldots$" denote higher order magnetic / electric multipole moments~\cite{MukamelPoNLS}. Here  $C_{\alpha}$ denotes the charge of the $\alpha$th particle in the system.  
When the sum over all charges is performed, the first two terms are the electric transition dipole moment $\boldsymbol\mu_{\ell a}$ and quadrapole $Q^{\nu_1,\nu_2}_{\ell a}$ moment (contracted over $\mathbf{k}$). The only term explicitly written term in the second bracket is the magnetic dipole moment $\mathbf{m}_{\ell a}$.

\section{Model two-chromophore system \label{SI:param}}
We consider a two-chromophore system with Hamiltonian
\begin{align}
\hat{H} = &\ket{1}\bra{1}E_0 + \ket{2}\bra{2}(E_0 + \Delta E) + V \ket{1}\bra{2} +\nonumber \\ 
&\ket{1,2}\bra{1,2}(2E_0 + \Delta E) + h.c. \;,
\end{align}
with $E_0 = 12038$cm$^{-1}$, $\Delta E= 323.61$cm$^{-1}$ and $V=117.56$cm$^{-1}$ and $\ket{k}$ representing the system with chromophore $k$ in the excited state and $\ket{k,k'}$ the two chromophore double excited states.  Strong electronic interaction $V$ between the chromophores lead to the formation of delocalized exciton eigenstates of the form:
\begin{equation}
\ket{\xi} = \sum_k c_k^{(\xi)} \ket{k} \;,
\end{equation}
with associated eigenenergies $E_{\pm} = 12200 \pm 200 $cm$^{-1}$, and coefficients $c_{k}^{-} = \{-\sin(2\pi/5),\cos(2\pi/5) \}$,   $c_{k}^{+} = \{\cos(2\pi/5),\sin(2\pi/5)\}$.  
The dipole moments are taken to be $\muB_1 = d[1, 0.5, 0]$ and $\muB_1 = d [0, 1, 0]$ where $d$ is a constant. The specific value of  $d$ is irrelevant  as we considering only excitonic CD effects, which have the same proportionality to $d$ as the achiral signal that we are normalising to unity via rescaling. The chromophores are separated by a distance $\Delta R =  [0, 1, -1]$nm, which results in an effective magnetic dipole moment
\begin{align}
\tilde{m}_{\xi}  &= i( C_1^{(\xi)} \muB_1 \times \mathbf{R}_{1}+C_2^{(\xi)} \muB_2 \times \mathbf{R}_{2})  \nonumber \\ 
&= i( C_1^{(\xi)} \muB_1 \times \Delta R - C_2^{(\xi)} \muB_2 \times \Delta R)/2 \;.
\end{align}

We consider a simple decoherence in the form of pure dephasing. The associated Lindblad operators are defined via $L_{C}(\rho)=C\rho C^\dagger -\left(C^\dagger C \rho +\rho C^\dagger C\right)/2$. The excited state dynamics of this system is described by the master equation 
\begin{equation}
\frac{\partial}{\partial t}  \rho(t) = -i[\hat{H},\rho(t)] + \gamma( \mathcal{L}_{\sigma_1}(\rho) + \mathcal{L}_{\sigma_2}(\rho) ) \;,
\end{equation}
where $\sigma_k = \ket{k}\bra{k} + \sum_{k' \neq k}\ket{k,k'}\bra{k,k'}$ and the dephasing rate $\gamma = 54.8~\textrm{cm}^{-1}$.

\section{Principles of two-dimensional chiral spectroscopy in brief \label{SI:2DES}}

Two dimensional spectroscopy involves excitation of medium with three coherent pulses, each of which have different wavevectors as discussed in the main text. These pulses are time ordered (no significant overlap in electric fields) with controlled delays between the centres; these delays are varied in each run of the experiment and the third order signal is measured at a particular phase matched direction via heterodyne detection with an electric field $\mathbf{E}_{\rm LO}(\mathbf{r},t)$.  Within the dipole approximation, we can write this signal in terms of a third order response function $\tensor{S}(t_3,\ldots)$ and a time varying electric field from the pulses $\mathbf{E}(\mathbf{r},t)$
\begin{align}
\mathbf{P}(\mathbf{r},t) =   &\iint d\mathbf{r_3}  \int_{-\infty}^{\infty} dt_3  \iint d\mathbf{r_2}  \int_{-\infty}^{\infty} dt_2  \iint d\mathbf{r_1}  \int_{-\infty}^{\infty} dt_1 \nonumber \\
&\mathbf{E}(\mathbf{r}_3,t-t_3) \mathbf{E}(\mathbf{r}_2,t-t_3-t_2) \mathbf{E}(\mathbf{r}_1,t-t_3-t_2-t_1)  \vdots  \nonumber \\ 
&\tensor{S}(t_3,t_2,t_1; \mathbf{r},\mathbf{r}_3,\mathbf{r}_2,\mathbf{r}_1) \;.
\end{align}

Assuming the interaction region in our medium is much longer than an optical wavelength, the induced signal field grows linearly (assuming the back-reaction on the pulses can be neglected) with the sample, in directions which satisfy phase matching conditions. A path through the sample $\tilde{\mathbf{r}}$ satisfies a phase matching condition if $(n_1 \mathbf{k}_1 +n_2 \mathbf{k}_2 +n_3 \mathbf{k}_3) \cdot \tilde{\mathbf{r}} \ll  \pi$ for $n_j$ integers; $n_1 = -1$, $n_2 = 1$, $n_3 = 1$ corresponds to the "rephasing" direction shown in the paper.  Therefore our signal field $\mathbf{E}_s \propto i \omega_s P_s(t)$ is related to $P_s(t)$, the Fourier component of the polarization in the direction we are measuring, assuming we have perfect phase matching.  We detect the intensity $|\mathbf{E}_s + \mathbf{E}_LO|^2$ and compare it to $|\mathbf{E}_LO|^2$  along to measure $\mathbf{E}_s$.  The total signal is then given by
\begin{align}
\mathrm{Sig} = 2\omega_s \mathrm{Re} \; \iint d^3\mathbf{r}  \int_{-\infty}^{\infty} dt \mathbf{E}_{\rm LO}(\mathbf{r},t) \cdot \mathbf{P}(\mathbf{r},t) \;.
\end{align}
For an ideal experiment we eventually some component (determined by polarization choices) of the nonlinear response tensor, Fourier transformed over the first and last time variables $S(\omega_3,\tau,\omega_1) = \int_0^{\infty} dt_1 \int_0^{\infty} dt_3 S(t_3,\tau,t_1)$.  

Similar expressions can be derived included chirality, in which one of the interactions can be with an interaction with the magnetic fields or the gradient of the electric field.  For our example which contains only excitonic CD, it is sufficient to use our example but include the positions of the chromophores within the complex (c.f.~ for example~\cite{Holdaway2016,Abramavicius2006}).  Including intrinsic magnetic dipole moments and electric quadrupole moments can also be achieved~\cite{Abramavicius2006-2}.

For geometries such as "GRAPES" the time delays between the pulses will vary along the phase matching direction for the rephasing signal (which BOXCARS explicitly avoids), leading to a non-trivial relation between the signal electric field and Polarization.  By measuring the variation of the signal field (via Heterodyne detection) in space it is possible to collect information about multiple delay times (with the exception of the delay between the second and third pulses) in a single shot.

\section{Impact of optical rotation within the medium \label{SI:CwsnT}}
As our sample is chiral, our electric field polarizations will vary depending on the distance propagated through the sample, and change for different frequency. Such changes will mean the pulses no longer strictly obey the orthogonality conditions we require to eliminate achiral signals.  Therefore we should determine whether this effect is significant in an experimental configuration.  

The polarization of an electric field within a medium is determined by the coupled Maxwell-Liouville equations, see for example~\cite{MukamelPoNLS}.  For a field which varies only along $z$ we have
\begin{equation}
 \pdsq{z} \colvec{E_x(z,t)}{E_y(z,t)} = \frac{1}{c^2} \pdsq{t}  \colvec{E_x(z,t)+4\pi P_x(z,t)}{E_y(z,t)+4\pi P_y(z,t)} \;.
\label{eq:Maxwell}
\end{equation}
We are interested in the forward propagating field in the direction of $\mathbf{k}$, so we write this component as $E_x(z,t) = En_x(z,t) \exp(i[k z - \omega_0 t])$ and the same for the polarization. If we are only expanding our polarization only to first order (sufficient to capture the effects we are interested in) our polarization can also be expressed in Fourier space as $\tilde{P}_a(z,\omega) = S_{ab}(\omega) E_{b}(z,\omega)$.  
This polarization will have the same forward exponential factor and can also be expressed in the envelope form. We can therefore Fourier transform~\refeq{eq:Maxwell} (denoting $\tilde{En}_x(z,\omega)$ as the Fourier transform of $En_x(z,t)$ etc) and solve purely in terms of envelope terms
\begin{align}
 ik\pd{z} \colvec{\tilde{En}_x(z,\omega-\omega_0)}{\tilde{En}_y(z,\omega-\omega_0)} = &\left(k^2 - \frac{\omega^2}{c^2} \right) \colvec{\tilde{En}_x(z,\omega-\omega_0)}{\tilde{En}_y(z,\omega-\omega_0)} \nonumber \\
 &-  \frac{4 \pi\omega^2}{c^2}\colvec{\tilde{Pn}_x(z,\omega-\omega_0)}{\tilde{Pn}_y(z,\omega-\omega_0)} \;.
\label{eq:Maxwell2}
\end{align}
We have neglected the second $z$ derivative of the electric field in the equation above, as significant change is not expected over the order of a single wavelength.  Factoring out the real component of the first order polarization $\mathbf{Re}(S^{(1)}_{xx})(\omega) = \mathbf{Re}(S^{(1)}_{yy})(\omega)= \sqrt{(n(\omega)-1)/4\pi}$ (which we assume to be almost entirely due to the solvent, which is assumed to not be birefringent) to give the refractive index, hence we have $k = \omega n(\omega)/c$ and this component cancels. The envelope in the direction $a$ obeys the differential equation
\begin{equation}
 \pd{z} En_a(z,\omega-\omega_0) \approx -\frac{2 \pi \omega}{n(\omega) c} [\mathrm{Im}(Pn_a^{(1)})(z,\omega-\omega_0)] \;.
\label{eq:env_z}
\end{equation}
In terms of the tensor elements of the first order (frequency space) response function $S^{(1)}_{ab}$ we have
\begin{align}
 \pd{z}  \colvec{En_x(z,\omega-\omega_0)}{En_y(z,\omega-\omega_0)} \approx &\frac{2 \pi \omega}{n(\omega) c} \left( \begin{array}{cc}
-\mathrm{Im}[S^{(1)}_{xx}(\omega)] & i S^{(1)}_{xy}(\omega) \\
i S^{(1)}_{yx}(\omega) & -\mathrm{Im}[S^{(1)}_{yy}(\omega)]  \end{array} \right) \nonumber \\ 
&\colvec{ En_x(z,\omega-\omega_0)}{ En_y(z,\omega-\omega_0)} \;.
\label{eq:env_z2}
\end{align}
We then note $S^{(1)}_{yx} = -S^{(1)}_{xy}$, as these terms will be identical besides switching two of the indices on a tensor average of odd rank (e.g.~$\langle Q^{\nu_1,\nu_2} \mu^{\nu_3}\rangle$), we obtain a solution of the form
\begin{align}
\colvec{En_x(z,\omega-\omega_0)}{En_y(z,\omega-\omega_0)} \approx e^{-\frac{\alpha z}{2}}  &\left( \begin{array}{cc}
\cosh\left(\frac{\beta z}{2}\right) & -i\sinh\left(\frac{\beta z}{2}\right) \\
i\sinh\left(\frac{\beta z}{2}\right) & \cosh\left(\frac{\beta z}{2}\right) \end{array} \right) \nonumber \\ 
&\colvec{ En_x(0,\omega-\omega_0)}{ En_y(0,\omega-\omega_0)} \;,
\label{eq:sample_effect}
\end{align}
where the parameters which are constant w.r.t $z$ are defined by
\begin{subequations} 
\begin{align}
 \alpha(\omega) &= -\rho \frac{4 \pi \omega}{n(\omega) c} \mathrm{Im} \left[S^{(1)}_{xx}(\omega) \right] \\
 \beta(\omega) &= \rho \frac{4 \pi i \omega}{n(\omega) c} S^{(1)}_{xy}(\omega)  \;.
\label{eq:alpha_beta}
\end{align}
\end{subequations}
Here $\rho$ the density of optically active molecules by volume. The matrix in~\refeq{eq:sample_effect} is commonly referred to as a Jones matrix, and full describes the linear response of the system exhibiting circular dichroism and optical rotation (but not birefringence or linear dichroism). The factors that drop out are $\beta = \eta/2+i\delta$ where $\eta = \alpha_L - \alpha_R$ is the circular dichroism and $\delta = |k| (n_L - n_R)$ is the optical rotation and $\alpha = (\alpha_L + \alpha_R)/2$ is the mean absorption.  To see this we consider the impact on left / right circularly polarized light ($E_x(0)= 1$ and $E_y(0) = \pm i$).  In this case we have $\mathbf{E}_{L/R}(z) = \mathbf{E}_{L/R}(0)\exp[-(\alpha \pm \beta) z/2]$ from~\refeq{eq:sample_effect}. The ratio of the intensities of the two fields for the each circular polarization direction is therefore $I_L / I_R = \exp[-\eta z]$.  

We can extract the polarization from \refeq{eq:sample_effect} by dividing the field by its amplitude. For our pulses, we have to change $z$ to the direction of propagation. In a truly colinear geometry, when all pulses have identical carrier frequencies, the optical rotation will not change the angles between polarization since they will all rotate equally.  However, for the BOXCARS and GRAPES configuration we will have more impact because the rotation will allow some of the non-chiral third order response function to contribute.  Despite the dependence on the non-chiral response function, this signal is still dependent on chirality (otherwise there would be no optical rotation). To guarantee this effect is weak we require $\mathrm{Im}[\beta(\omega)] L \ll 1/M$, where $M$ is the relative amplitude of the non-chiral and chiral signals. The actual conditions will depend on the angles between the wavevectors, allowing more tolerance.  
To quantify this we look at two the polarization of two rays of identical frequency light traveling through our sample. One initially polarized along $\polB_1(0) = \hat{y}$ and traveling along $\hat{z}$ the other polarized along $\polB_2(0)=\hat{x}$ and traveling along $\hat{z} \cos(\theta) + \hat{y} \sin(\theta)$.  As the rays enter a sample we have $\polB_1(0) \cdot \polB^*_2(0) =0$, however after traveling a distance $L$ through the sample we have
\begin{align}
\polB_1(L) \cdot \polB^*_2(L) \propto -i[&\sinh(\beta L/2)\cosh(\beta^* L/2) \nonumber \\
&+ \cos(\theta)\sinh(\beta^* L/2)\cosh(\beta L/2)] \;.
\label{eq:angle_L}
\end{align}
The proportional to is used as the propagation formula included decay as well as rotation.  For simplicity we ignore all circular dichroism (giving $\beta =i\delta$) and thus obtain
\begin{equation}
\polB_1(L) \cdot \polB^*_2(L) = [1- \cos(\theta)] \sin(\delta L)/2 \;.
\label{eq:angle_L2}
\end{equation}
Using \refeq{eq:angle_L2} we can refine our previous estimate for the requirements on the interaction depth to $\delta L \ll 2/M[1- \cos(\theta)]$, or if $\theta \sim 0$ we can write $\delta L \ll 4 /M\theta^2$.  We expect $\delta M$ to be of a similar order to the absorption of the sample anyway, so this condition is actually weaker when $\theta$ is small, than the condition for an optically thin sample, which we require for performing 2DS anyway. Therefore the impact of optical rotation within the medium can likely be neglected. 
\end{appendices}

\end{document}